\documentclass[12pt]{iopart}
\usepackage{iopams}
\usepackage{cite}
\usepackage{graphicx}
\newcommand{\tfrac}[2]{\mbox {${\textstyle \frac{ #1 }{ #2 }}$}}
\newcommand{\bra}[1]{\left\langle #1 \right |}
\newcommand{\ket}[1]{\left | #1 \right\rangle}

\begin{document}
\title{Three-tangle for mixtures of generalized GHZ and generalized W states}
\author{C\ Eltschka$^1$, A\ Osterloh$^2$, J\ Siewert$^1$,
        and A\ Uhlmann$^3$}
\address{$^1$ Institut f\"ur Theoretische Physik, Universit\"at 
              Regensburg, D-93040 Regensburg, Germany}
\address{$^2$ Institut f\"ur Theoretische Physik, Leibniz Universit\"at
              Hannover, D-30167 Hannover, Germany}
\address{$^3$ Institut f\"ur Theoretische Physik,
         Universit\"at Leipzig, D-04109 Leipzig, Germany}
       \ead{christopher.eltschka@physik.uni-regensburg.de}
\begin{abstract}
  We give a complete solution for the three-tangle of mixed
  three-qubit states composed of a generalized GHZ state,
  $a\left|000\right>+b\left|111\right>$, and a generalized W state,
  $c\left|001\right>+d\left|010\right>+f\left|100\right>$. Using the
  methods introduced by Lohmayer {\em et al.} we provide explicit
  expressions for the mixed-state three-tangle and the corresponding
  optimal decompositions for this more general case. Moreover, as a
  special case we obtain a general solution for a family of states
  consisting of a generalized GHZ state and an orthogonal product state.
\end{abstract}
\section{Introduction}
The occurrence of entanglement in multipartite systems
is one of the most important and distinctive features in
quantum theory~\cite{PlenioReview2007,HorodeckiRMP2007}. 
With the ever-increasing number of applications of entanglement, 
its quantification has become one of the foremost topics in 
contemporary quantum information research.

While entanglement of pure and mixed states of two qubits is already
well understood~\cite{Werner1989, Bennett1996, BennettDiVincenzo96-a,
  Wootters1998,Wootters2001}, to date there is no generally accepted
theory for classification and quantification of entanglement in
multipartite qubit systems. For three-qubit systems, numerous
interesting results have been found~\cite{Duer1999,
  Coffman2000, Duer00-a, Acin2000, Carteret2000, Acin01,
  AcinBruss2001, Wei2003, Levay2005, Yu2006, LOSU2006,
  Loss3qubit2007}. A complete SLOCC characterization of three-qubit
entanglement has been achieved only for pure states~\cite{Coffman2000,
  Duer00-a}. It leads to a schematic characterization for mixed
states~\cite{AcinBruss2001}. A crucial concept for this is the
so-called three-tangle, a polynomial invariant for three-qubit states
that quantifies the three-partite entanglement contained in a pure
three-qubit state (the three-tangle is equal to the modulus of the
hyperdeterminant~\cite{Caley1845,Miyake2003}). However, even for the
simplest case of rank-2 mixed states, no general expression is known
for its three-tangle.

Recently, Lohmayer {\em et al.}~\cite{LOSU2006} have provided an
analytic quantification of the three-tangle for a representative
family of rank-2 three-qubit states, namely for mixtures of a
symmetric GHZ state and an orthogonal symmetric W state. In this
article we show that by applying the methods
of~\cite{LOSU2006,OSU2007} these results can be extended to rank-2
mixtures of a generalized GHZ state and an orthogonal generalized W state. This
article is organized as follows. In Section 2, we introduce some basic
terminology and give a precise formulation of the problem whose
general solution we outline in Section 3. In Section 4 we discuss
special cases of this solution, in particular we find the three-tangle
for rank-2 mixtures of generalized GHZ states and certain orthogonal product
states.

\section{Notations and formulation of the problem}

Consider the state $|\psi\rangle$ in a three-qubit Hilbert space
$|\psi\rangle \in {\cal H}_A\otimes{\cal H}_B\otimes{\cal H}_C$.
Its coefficients with respect to a basis of product states 
(the `computational basis') are
$\psi_{jkl}=\langle jkl|\psi\rangle$, $j,k,l\in \{0, 1\}$.
An important measure for the entanglement in pure three-qubit states is
the three-tangle (or residual tangle) introduced in~\cite{Coffman2000}.
The three-tangle of $|\psi\rangle$ is a so-called 
polynomial invariant~\cite{VerstraeteDM03,Leifer04}
and  can be written in terms of the coefficients $\psi_{ijk}$ as
\begin{eqnarray}
  \label{eq:3tangle}
  \tau_3(\psi)\  &=&\  4|d_1-2d_2+4d_3|\\
  d_1\  &=&\  \psi_{000}^2\psi_{111}^2 + \psi_{001}^2\psi_{110}^2 + \psi_{010}^2\psi_{101}^2
  + \psi_{100}^2\psi_{011}^2\nonumber\\
  d_2\  &=&\  \psi_{000}\psi_{111}\psi_{011}\psi_{100} +
  \psi_{000}\psi_{111}\psi_{101}\psi_{010} + \psi_{000}\psi_{111}\psi_{110}\psi_{001} \nonumber\\
  && {}+\psi_{011}\psi_{100}\psi_{101}\psi_{010} + \psi_{011}\psi_{100}\psi_{110}\psi_{001} +
  \psi_{101}\psi_{010}\psi_{110}\psi_{001}\nonumber\\
  d_3\  &=&\  \psi_{000}\psi_{110}\psi_{101}\psi_{011} + \psi_{111}\psi_{001}\psi_{010}\psi_{100}\ \ .\nonumber
\end{eqnarray}
The three-tangle of a mixed state
\begin{equation}
  \label{eq:mixed}
  \rho=\sum_jp_j\pi_j,\qquad 
        \pi_j=\frac{\left|\phi_j\rangle\!\langle\phi_j\right|}
                   {\langle\phi_j|\phi_j\rangle}
\end{equation}
can be defined as convex-roof extension~\cite{Uhlmann00-a} of the pure
state three-tangle,
\begin{equation}
  \label{eq:convroof}
  \tau_3(\rho)=\min_{\mathrm{decompositions}}\sum_jp_j\tau_3(\pi_j).
\end{equation}
A given decomposition $\{q_k, \pi_k: \rho=\sum_k q_k\pi_k\}$ with 
$\tau_3(\rho)=\sum_k q_k \tau_3(\pi_k)$ is called {\em optimal}.
We note that $\tau_3(\rho)$ is a convex function on the convex 
(and compact) set $\Omega$ of
density matrices $\rho$.

In this paper, we determine three-tangle and optimal decompositions
for the family of mixed three-qubit states
\begin{equation}
  \label{eq:rho(p)}
  \rho(p) = p\left|gGHZ_{a,b}\right>\!\left<gGHZ_{a,b}\right| +
  (1-p)\left|gW_{c,d,f}\right>\!\left<gW_{c,d,f}\right|
\end{equation}
composed of a generalized GHZ state
\begin{equation}
  \label{eq:gGHZ}
  \left|gGHZ_{a,b}\right> = a\left|000\right> + b\left|111\right>,\qquad |a|^2+|b|^2=1
\end{equation}
and a generalized W state
\begin{equation}
  \label{eq:gW}
  \left|gW_{c,d,f}\right> = c\left|001\right> + d\left|010\right> + f\left|100\right>,
  \qquad |c|^2+|d|^2+|f|^2=1\ \ .
\end{equation}
We note that $\tau_3(gW_{c,d,f})=0$ and
$\tau_3^{\mathrm{gGHZ}}:=\tau_3(gGHZ_{a,b})=4|a^2b^2|$. For the symmetric GHZ and
W state ($a=b=1/\sqrt{2}$ and $c=d=f=1/\sqrt{3}$) the problem and
results of \cite{LOSU2006} are recovered.

\section{The generic case}
In this section it is assumed that none of the coefficients is zero,
i.e. $a,b,c,d,f\neq0$. The opposite case corresponds to either a rank-2
mixture of a generalized GHZ and a biseparable state, or to a mixture of a
generalized W and a completely factorized state and will be studied in the
next section.

In the following, we will apply the methods developed
in~\cite{LOSU2006,OSU2007}. There it was shown that in order to find
the convex roof of an entanglement measure for rank-2 mixed states it
is useful to study the pure states that are superpositions of the
eigenstates of $\rho$
\begin{equation}
  \label{eq:superposition}
  \left|p,\varphi\right> =
  \sqrt{p}\left|gGHZ_{a,b}\right> - \sqrt{1-p}\;\rme^{\rmi\varphi}\left|gW_{c,d,f}\right>
\ \ .
\end{equation}
The three-tangle of these states is
\begin{equation}
  \label{eq:pure3tangle}
  \tau_3(p,\varphi) =
  4\left|p^2a^2b^2 - 4\sqrt{p(1-p)^3}\;\rme^{3\rmi\varphi}bcdf\right|\ \ .
\label{3tang-coeffs} 
\end{equation}
The phases of the coefficients in $\left|gGHZ_{a,b}\right>$ and
$\left|gW_{c,d,f}\right>$ merely produce different offsets for the relative
phase $\varphi$ in the expression for the three-tangle,
Eq.~(\ref{3tang-coeffs}). Therefore it suffices to consider the case
where all coefficients are positive real numbers.

In the following, it will be beneficial to introduce the definition
\begin{equation}
  \label{eq:s}
  s = \frac{4cdf}{a^2b} > 0\ \ .
\end{equation}
If we factor out the three-tangle $\tau_3^{\mathrm{gGHZ}}$ of the generalized
GHZ state, the three-tangle of the superposition
(\ref{eq:superposition}) can be written as
\begin{equation}
  \label{eq:pure3tangle-s}
  \tau_3(p,\varphi) =  \tau_3^{\mathrm{gGHZ}}\left|p^2-\sqrt{p(1-p)^3}\;\rme^{3\rmi\varphi}s\right|\ \ .
\end{equation}
Since $\tau_3^{\mathrm{gGHZ}}$ is just a constant factor, the behaviour
of this function of $p$ and $\varphi$
is completely determined by the value of the parameter $s$.

As a first step, we identify the {\em zero-simplex} containing all
mixed states $\rho(p)$ with $\tau_3(\rho(p))=0$. Its corner states are obtained
as the zeros of Eq.~(\ref{eq:pure3tangle-s}). One obvious solution is
$p=0$, which corresponds to a pure generalized W state. Therefore, in the
calculation of the other solutions we can assume $p>0$ and the zeros
are determined by
\begin{equation}
  \label{eq:reduced}
  \sqrt{p^3}=\sqrt{(1-p)^3}\rme^{3\rmi\varphi}s\ \ .
\end{equation}
Since $p$ and $s$ are real and positive, this implies\footnote{Note
  that the $2\pi/3$-periodicity is due to the fact that this relative
  phase is induced by the local transformation ${\rm diag}\{\exp(\rmi
  2\pi/3),1\}$ on each qubit.}
\begin{equation}
  \label{eq:phi}
  \varphi=n\frac{2\pi}{3},\qquad n\in\mathbb{N} \ \ .
\end{equation}
For $p$, we then get the solution
\begin{equation}
  \label{eq:p0}
  p_0=\frac{s^{2/3}}{1+s^{2/3}}
  =\frac{\sqrt[3]{16c^2d^2f^2}}{\sqrt[3]{a^4b^2}+\sqrt[3]{16c^2d^2f^2}} \ \ .
\end{equation}
This means that in addition to the state $\ket{gW_{c,d,f}}$ the three-tangle
vanishes for $\left|p_0,n\cdot 2\pi/3\right>$, $n=0,1,2$. All mixed states
whose density matrices are convex combinations of those four states
have zero three-tangle. On the Bloch sphere with gGHZ and gW at its
poles, this corresponds to a simplex with those four states at the
corners. All $\rho(p)$ with $p<p_0$ are inside this set, and therefore
$\tau_3(\rho(p))=0$ for $0\leq p\leq p_0$.

In order to determine the mixed three-tangle of $\rho(p)$ for $p>p_0$, we
note that for any fixed $p$, $\tau_3(p,\varphi)$ takes a minimum at $\varphi_0=0$
which due to the symmetry of $\tau_3$ is repeated at $\varphi_1=2\pi/3$ and
$\varphi_2=4\pi/3$. Consequently, for any value of $p$ the state $\rho(p)$ can be
decomposed into the three states $\ket{p,\varphi_i}$, $i=0,1,2$. Therefore
the characteristic curve $\tau_3(p,0)$ is an upper bound to $\tau_3(\rho(p))$.
Moreover is it known to give the correct values for the three-tangle
at $p=p_0$ (at the top face of the zero simplex) and $p=1$
($\rho(1)=\ket{gGHZ_{a,b}}\!\bra{gGHZ_{a,b}}$). However, if there is a range of
values where $\tau_3(p,0)$ is a concave function, there are
decompositions for $\rho(p)$ with a lower average
three-tangle~\cite{LOSU2006}. Therefore it is important to examine
where the function $\tau_3(p,0)$ is concave for $p\geq p_0$.

For $\varphi=0$ and $p\geq p_0$, the term inside the absolute value bars in
(\ref{eq:pure3tangle-s}) is real and positive, and the characteristic
curve $\tau_3(p,0)$ is equal to
\begin{equation}
  \label{eq:t(p)}
  t(p)\ =\ \tau_3^{\mathrm{gGHZ}}\cdot(p^2 - \sqrt{p(1-p)^3}s)\ \ .
\end{equation}
Concavity of $t(p)$ is indicated by a negative sign of its second
derivative
\begin{equation}
  \label{eq:t''(p)}
  t''(p)\ =\ \tau_3^{\mathrm{gGHZ}}\left(2-
      \frac{8p^2-4p-1}{4p\sqrt{p(1-p)}}s\right)\ \ .
\end{equation}
The limit $p\to 1$ ($p=1-\varepsilon$) in (\ref{eq:t''(p)}) gives
\begin{equation}
  \label{eq:p->1}
  t''(1-\varepsilon)\ =\ -\frac{3\tau_3^{\mathrm{gGHZ}}s}{4\sqrt{\varepsilon}}
               + 2\tau_3^{\mathrm{gGHZ}} + \Or(\varepsilon^{1/2})\ \ ,
\end{equation}
that is, $t(p)$ is concave close to $p=1$. On the other hand, for
small $p$
\begin{equation}
  \label{eq:p->0}
  t''(p)=\frac{\tau_3^{\mathrm{gGHZ}}s}{4p^{3/2}} + \Or(p^{-1/2})\ \ .
\end{equation}
That is, close to $p=0$ we find that $t(p)$ 
is convex (note that due to the absolute
value, $\tau_3(p,0)$ is actually concave close to
$p=0$). Due to continuity, there must be at least one zero of $t''(p)$
in between. Moreover we note that the third derivative
\begin{equation}
  \label{eq:t'''(p)}
  t'''(p)=\frac{-3\tau_3^{\mathrm{gGHZ}}s}{8p^2\sqrt{p(1-p)^3}}\leq 0
\end{equation}
is negative for all values of $p$. Thus $t''(p)$ is strictly
monotonous and has precisely one zero, implying that $t(p)$ is convex
before and concave after that point. As the mixed state three-tangle
is convex, the characteristic curve needs to be convexified where it
is concave in the interval $[p_0,1]$. Since the concavity extends up
to $p=1$, corresponding to the state $\ket{gGHZ_{a,b}}$, that state has to
be part of the optimal decomposition~\cite{OSU2007} in this interval.

The symmetry and the results in~\cite{LOSU2006} suggest that a good
ansatz for the optimal decomposition is
\begin{equation}
  \label{eq:decomp}
  \rho(p) = \alpha\left|gGHZ_{a,b}\right>\!\left<gGHZ_{a,b}\right|
  + \frac{1-\alpha}{3}
  \sum_{k=0}^2\left|p_1,k\cdot \tfrac{2\pi}{3}\right>\!\left<p_1,k
               \cdot \tfrac{2\pi}{3}\right|
\end{equation}
where $p_1$ is chosen such that the mixed-state three-tangle becomes
minimal. The value of $\alpha$ is fixed by $p$ and $p_1$:
\begin{equation}
  \label{eq:alpha}
  \alpha = \frac{p-p_1}{1-p_1}\ \ .
\end{equation}
The average three-tangle for this decomposition is ($p>p_0$)
\begin{equation}
  \label{eq:mixed3tangle}
  \tau_3^{\mathrm{conv}}(p,p_1)\ =
  \ \frac{p-p_1}{1-p_1}\cdot\tau_3^{\mathrm{gGHZ}}
  + \frac{1-p}{1-p_1}\cdot t(p_1)\ \ .
\end{equation}
This describes a linear interpolation between $\tau_3(p_1,0)$ and
$\tau_3^{\mathrm{gGHZ}}$. Note that for $p<p_1$, (\ref{eq:decomp}) ceases
to be a valid decomposition because $\alpha$ becomes negative.

To find the minimum in $p_1$ for given $p$, we look for the zeros of
the derivative $\partial \tau_3^{\mathrm{conv}}/ \partial p_1$. The resulting equation
has the solution
\begin{equation}
  \label{eq:p1-solution}
  p_1^{\mathrm{noabs}}=\frac{1}{2}+\frac{1}{2\sqrt{1+s^2}}\ \ .
\end{equation}
Note that for $s>2\sqrt{2}$ we get $p_1^{\mathrm{noabs}}<p_0$. In that
case the minimum is reached at the border $p_1=p_0$ of the considered
interval $[p_0,1]$, and therefore
\begin{equation}
  \label{eq:p1}
  p_1=\max\{\, p_0,\frac{1}{2}+\frac{1}{2\sqrt{1+s^2}}\, \}\ \ .
\end{equation}
Putting it all together, we present the central result of this article
\begin{equation}
  \label{eq:combined}
  \tau_3(\rho(p))\ =\ 
  \cases{0 & for $0\leq p\leq p_0$\\
    \tau_3(p,0) & for $p_0\leq p\leq p_1$\\
    \tau_3^{\mathrm{conv}}(p,p_1) & for $p_1\leq p\leq 1$}
\end{equation}
where $p_0$ is given by (\ref{eq:p0}), $p_1$ by (\ref{eq:p1}),
$\tau_3(p,0)$ by (\ref{eq:pure3tangle}) and $\tau_3^{\mathrm{conv}}(p,p_1)$
by (\ref{eq:mixed3tangle}). The corresponding optimal decompositions
are
\begin{equation}
  \label{eq:decomps}
  \rho(p)=
  \cases{\frac{p}{p_0}\rho_\Delta(p_0)+\frac{p_0-p}{p_0}\pi_{\mathrm{gW}} & for $0\leq p\leq p_0$\\
    \rho_\Delta(p) & for $p_0\leq p\leq p_1$\\
    \frac{1-p}{1-p_1}\rho_\Delta(p_1) +
    \frac{p-p_1}{1-p_1}\pi_{\mathrm{gGHZ}} & for $p_1\leq p\leq 1$}
\end{equation}
where
\begin{equation}
  \label{eq:rhodelta}
  \rho_\Delta(p)=\frac{1}{3}\sum_{k=0}^2\left|p,k\cdot \tfrac{2\pi}{3}\right>\!
\left<p,k\cdot \tfrac{2\pi}{3}\right|
\end{equation}
and $\pi_j$ as defined in (\ref{eq:mixed})

The curve (\ref{eq:combined}) is convex, and for all $p$ and $\varphi$:
$\tau_3(\rho(p))\leq\tau_3(p,\varphi)$. Therefore it is a lower bound to the
three-tangle of $\rho(p)$. On the other hand, for each $p$ we have given
an explicit decomposition realizing this lower bound. Thus it
represents also an upper bound and hence coincides with the
three-tangle of $\rho(p)$.

\begin{figure}[t]
  \centering
  \begin{tabular}{cc}
    (a) & \includegraphics[width=0.8\textwidth]{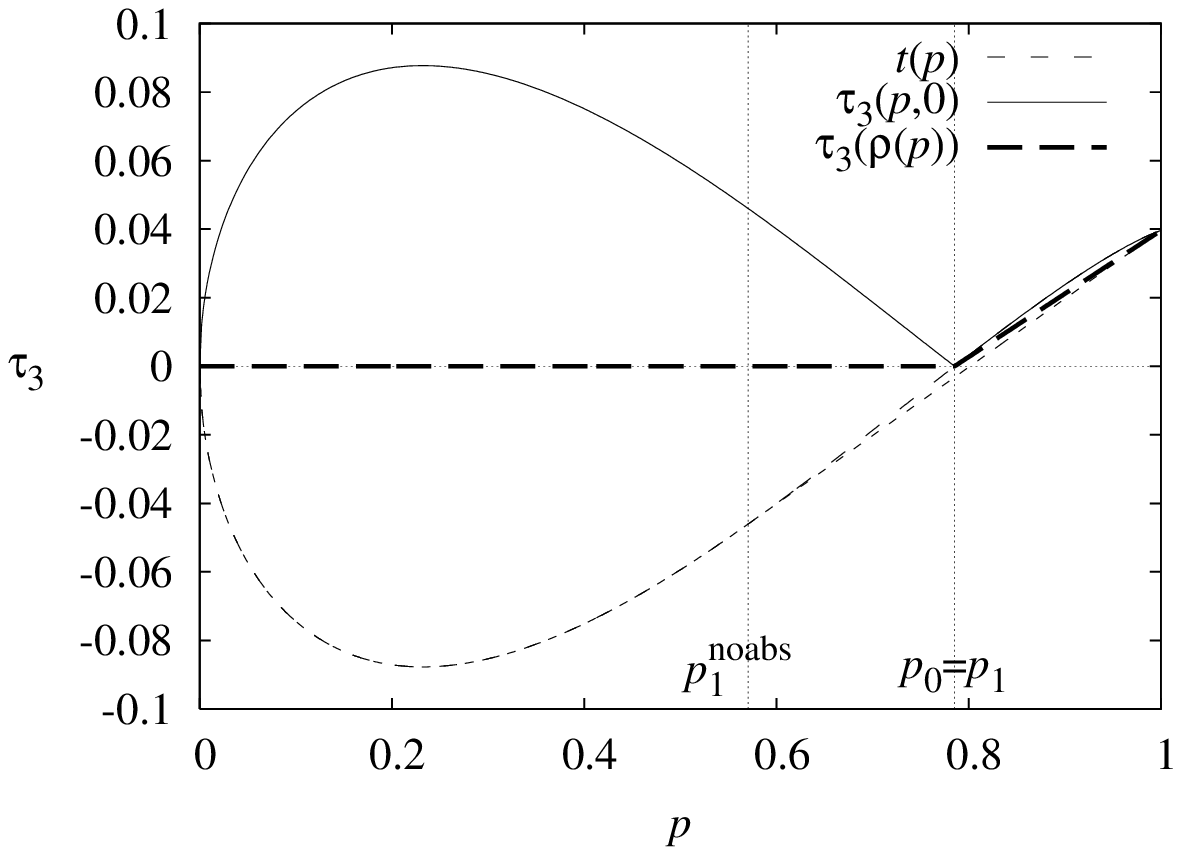}   \\
    (b) & \includegraphics[width=0.8\textwidth]{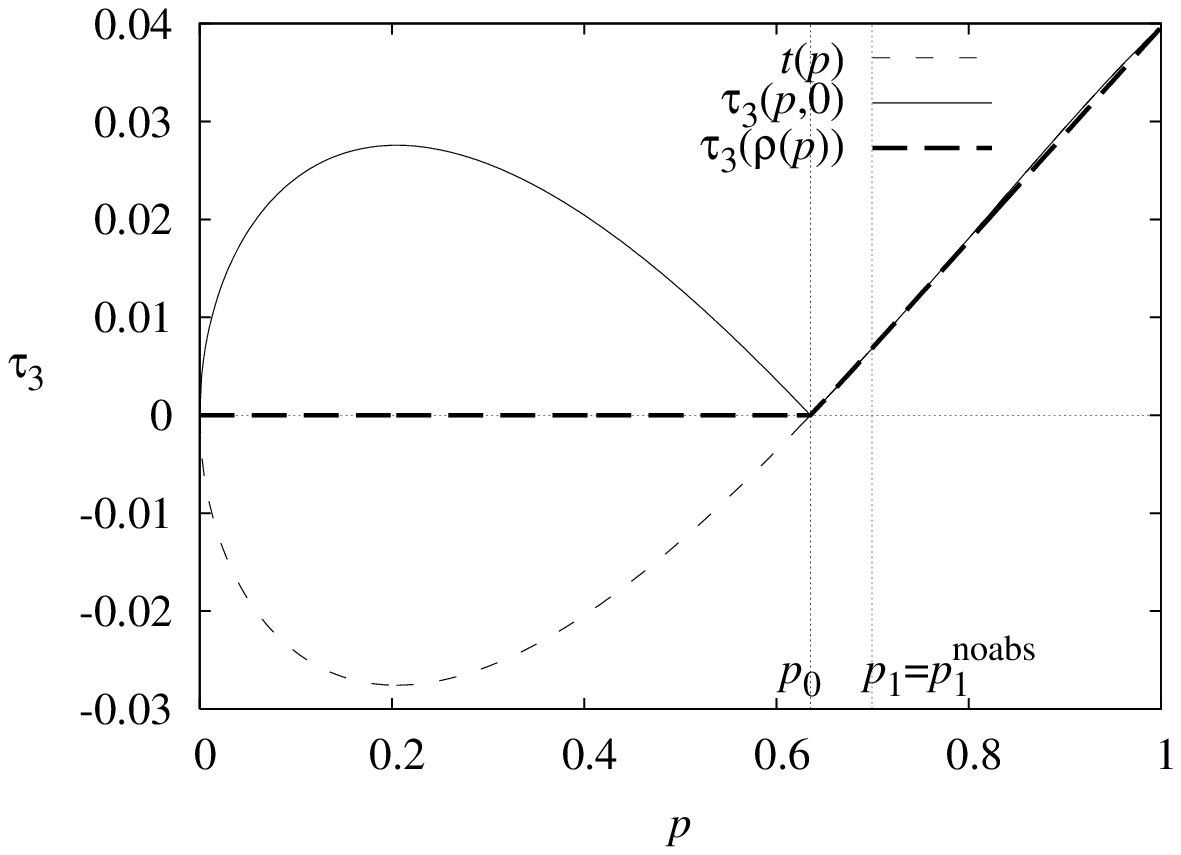}   
  \end{tabular}
  \caption{Three-tangle for $s=7>2\sqrt{2}$ (a) and $s=2.3<2\sqrt{2}$
    (b). In both cases $\tau_3^{\mathrm{gGHZ}}=0.0396$. The solid line is
    the minimal pure state tangle, $\tau_3(p,0)$
    (\ref{eq:pure3tangle-s}). The short-dashed line is $t(p)$
    (\ref{eq:t(p)}). The dotted vertical lines show the positions of
    $p_0$ (\ref{eq:p0}), $p_1^{\mathrm{noabs}}$ (\ref{eq:p1-solution})
    and $p_1$ (\ref{eq:p1}), and the thick dashed line gived the
    resulting mixed three-tangle $\tau_3(\rho(p))$ (\ref{eq:combined}). In
    addition, the first figure shows as dotted line the curve which
    would result from using $p_1^{\mathrm{noabs}}$ instead of $p_1$ in
    (\ref{eq:mixed3tangle}).}
  \label{fig:1}
\end{figure}

\section{Special cases}

In this section we will discuss various special cases of our
general solution (\ref{eq:combined}).

First, we briefly demonstrate that the results for the symmetric GHZ
state and the symmetric W state in~\cite{LOSU2006} are reproduced.
Indeed, the general behaviour described in Section 3 (that is,
analytic properties of the three-tangle, optimal decompositions)
matches the one found in~\cite{LOSU2006}, so we only have to check the
values of $p_0$ and $p_1$. In the symmetric case we have
$a=b=1/\sqrt{2}$ and $c=d=f=1/\sqrt{3}$, resulting in
\begin{equation}
  \label{eq:gwsymm}
  s = \frac{2^{7/2}}{3^{3/2}}\ \ .
\end{equation}
Inserting this in (\ref{eq:p0}) and (\ref{eq:p1}) leads to
\begin{eqnarray}
  \label{eq:p0p1symm}
  p_0 &=& \frac{2^{7/3}/3}{1+2^{7/3}/3}
  = \frac{4\sqrt[3]{2}}{3+4\sqrt[3]{2}}\\
  p_1 &=& \frac{1}{2}+\frac{1}{2\sqrt{1 + 2^7/3^3}}
  = \frac{1}{2}+\frac{3}{2}\sqrt{\frac{3}{155}}
\end{eqnarray}
as found in~\cite{LOSU2006}.

Next, we consider the limiting cases where at least one of the
coefficients is $0$. Those require extra care as the
calculations above have been done under the assumption of 
non-vanishing coefficients.
However, since we are dealing with continuous functions, one
should expect that the results still apply, although
possibly in a degenerate form.

The first case we consider is when the generalized GHZ state degenerates into
a pure three-party product state. This corresponds to the limit $s\to\infty$.
However note that at the same time $\tau_3^{\mathrm{gGHZ}}\to 0$ such that
(\ref{eq:pure3tangle-s}) remains regular. This can be seen by looking at
the explicit form (\ref{eq:pure3tangle}). 
It is clear that in this case $\tau_3(\rho(p))=0$ for all $p$.

There are two non-equivalent ways to achieve this. One possibility is
$b=0$ which reduces the generalized GHZ state to $\left|000\right>$. In this
case, the three-tangle (\ref{eq:pure3tangle}) vanishes for all
superpositions (\ref{eq:superposition}), and therefore also all mixed
states anywhere inside the Bloch sphere have vanishing three-tangle.

The other way to get $s\to\infty$ is $a=0$ where the generalized GHZ state is
reduced to $\left|111\right>$. While $\rho(p)$ as a mixture of product
and $gW$ state again has no three-tangle, unlike in the case $b=0$ the
three-tangle does not vanish everywhere on the Bloch sphere. Equation
(\ref{eq:pure3tangle}) reduces to
\begin{equation}
  \label{eq:3tangle-g=0}
  \tau_3(p,\varphi)=16\sqrt{p(1-p)^3}\;cdf,
\end{equation}
which is independent of $\varphi$ and concave for all $p\in[0,1]$. Thus the
zero simplex degenerates into a {\em zero axis}. As long as $cdf>0$,
outside of this axis the three-tangle never vanishes. If both $a=0$
and $cdf=0$, the three-tangle is zero everywhere inside the Bloch
sphere.

The opposite limiting case is $s=0$, that is, when at least one of the
coefficients $c$, $d$, $f$ vanishes. Note that for the three-tangle it
does \emph{not} matter whether only one of them vanishes, resulting in
a product of a single qubit state with a generalized Bell state, or two of
them, resulting in a product of three single-qubit states: In all
cases (\ref{eq:pure3tangle-s}) reduces to
\begin{equation}
  \label{eq:pure3tangle-w0}
  \tau_3(p,\varphi)\ =\  \tau_3^{\mathrm{gGHZ}}\cdot p^2\ \ ,
\end{equation}
which is convex for all $p\in[0,1]$; indeed, (\ref{eq:p0}) and
(\ref{eq:p1}) yield $p_0=0$ and $p_1=1$ at $s=0$. Consequently,
\begin{equation}
  \label{tau3mixedw=0}
  \tau_3(\rho(p))=\tau_3^{\mathrm{gGHZ}}\cdot p^2
\end{equation}
for all $p$. Even more, $\tau_3(\rho)=\tau_3^{\mathrm{gGHZ}} p^2$ for
\emph{any} mixed state $\rho$ inside the Bloch sphere with
$\bra{gGHZ_{a,b}}\rho\ket{gGHZ_{a,b}}=p$. We would like to point out that this result
reminds of the situation both for two-qubit
superpositions~\cite{Abouraddy2001} and for two-qubit mixtures of an
arbitrary entangled state and an orthogonal product state.

\section{Conclusion}
In this paper, we have given explicit expressions for the three-tangle
of mixtures $\rho(p)$ according to (\ref{eq:rho(p)}) of arbitrary generalized
GHZ and orthogonal generalized W states, including the limiting cases where
those states are reduced to product states. We have found that the
qualitative pattern described in~\cite{LOSU2006} for mixtures of
symmetric GHZ and W states holds also more generally. Up to a certain
value $p_0$ given by (\ref{eq:p0}), the mixed three-tangle vanishes.
The optimal decomposition for those states consists of the pure states
(\ref{eq:superposition}) for which the three-tangle is zero. One is
always the generalized W state at the bottom of the Bloch sphere. The other
three form an equilateral horizontal triangle at the height of $p_0$.
Note that those states do not depend on $p$ as long as $p\leq p_0$.

For $p> p_0$, there \emph{may} follow a region up to some value $p_1$
given by Eq.~(\ref{eq:p1}), where the mixed state three-tangle follows
the minimal pure state three-tangle (\ref{eq:pure3tangle-s}) with the
same value for $p$ (which for positive real coefficients is achieved
at $\varphi=0$). In this region, the optimal decomposition consists of the
three states with this property, which form a horizontal eqilateral
triangle with corners on the Bloch sphere and $\rho(p)$ in the center.
If $s\geq 2\sqrt{2}$, $p_1$ and $p_0$ coincide and this region with
``leaves'' of constant three-tangle in the convex roof
(cf.~\cite{LOSU2006}, Figure~2) is absent. This can be viewed as
contraction of this middle region into one point.

For $p>p_1$, the three-tangle grows linearly up to its maximum value
at $p=1$. The optimal decomposition in this case consists of the three
pure superposition states for $p=p_1$ with minimal three-tangle and
the generalized GHZ state. That is, the convex roof in the Bloch sphere is
affine for an entire simplex whose corners are given by the four pure
states that form the optimal decomposition. Moreover, we have
demonstrated how the results of this work connect to the findings for
the special case of mixtures of a symmetric GHZ and a symmetric W
state~\cite{LOSU2006}.

In principle, the scheme of three regions for $p$ values as outlined
above holds also in the limiting cases when some of the coefficients
in the states vanish, except that in this situation the ``outer
regions'' may shrink away. A common feature of these limits is a
$\varphi$-independent characteristic curve. If the generalized GHZ state
degenerates into a product state, $\tau_3(\rho(p))=0$ for all $p$. On the
other hand, for $s=0$ (i.e., at least one of the coefficients in the
generalized W states vanishes), both ``outer'' affine regions disappear and
the whole range of $p$ is covered by the ``middle region'' with a
strictly convex characteristic curve. This case corresponds to a
mixture of a generalized GHZ state and an orthogonal product state and the
exact convex roof of the three-tangle is obtained everywhere inside
the Bloch sphere.

\ack

We acknowledge interesting discussion with G\'eza T\'oth.
This work was supported by the Sonderforschungsbereich 631
of the German Research Foundation. JS.\ receives support from
the Heisenberg Programme of the German Research Foundation.

\section*{References}

\bibliographystyle{iopart-num}
\bibliography{Artikel_sGHZW.bib}
\end{document}